\documentclass{aa}
\usepackage{graphicx}
\usepackage{natbib}
\begin{document}
   \title{On the galactic chemical evolution of sulfur}


   \author{N. Ryde
          \inst{1,2}
      \and
          D. L. Lambert
      \inst{2}
       }

   \offprints{N. Ryde}

   \institute{ Uppsala Astronomical Observatory,
         Box 515,
         SE-751 20 Uppsala,
         Sweden\\
         \and Department of Astronomy,
         University of Texas,
         Austin, TX 78712-1083,
         USA\\
        \email{ryde@astro.uu.se}
         }

   \date{Received ; accepted }

   \abstract{Sulfur
abundances have been determined for ten
   stars
to resolve a debate in the literature on the Galactic chemical
   evolution of sulfur in the halo phase of the Milky Way.
Our analysis is based on observations of the S\,{\sc i} lines at
$9212.9$, $9228.1$, and  $9237.5$~\AA\ for stars for which the S
abundance was obtained previously from much weaker S\,{\sc i}
lines at $8694.0$ and $8694.6$~\AA . In contrast to the previous
results showing [S/Fe] to rise steadily with decreasing [Fe/H],
our results show that [S/Fe] is approximately constant for
metal-poor stars ([Fe/H]{\small
\raisebox{-0.02cm}{\begin{minipage}{0.2cm}
\raisebox{-0.2cm}{$< $} \\
\raisebox{0.08cm}{$\sim$ }
\end{minipage}}}$ -1$) at [S/Fe] $\simeq$ +0.3. Thus,
   sulfur behaves in a
   similar way to the other $\alpha$ elements, with an approximately constant [S/Fe] for
   metallicities lower than $\mathrm{[Fe/H]}\simeq -1$.
   We  suggest that the reason for the earlier claims of a
   rise of [S/Fe] is partly due to the use of the weak S\,{\sc i} $8694.0$ and $8694.6$\,\AA\
   lines and
   partly uncertainties in the determination of the metallicity
   when using Fe\,{\sc i} lines. The
   S\,{\sc i}~$9212.9$, $9228.1$, and  $9237.5$\,\AA\
   lines are preferred for an abundance analysis of sulfur for metal-poor stars.
}


   \maketitle
%

\section{Introduction}

Galactic chemical evolution of the $\alpha$ elements (O, Ne, Mg,
Si, S, Ar, and Ca) might be expected to be similar from element to
element. It is thought that their production in the early Galaxy
was dominated by synthesis in massive stars and ejection into the
interstellar medium by supernovae (SN Type II). Then, relative
abundances of $\alpha$ elements and of their abundances with
respect to iron reveal information about the yields from SN II and
the dependence of those yields on the initial metallicity.  After
a passage of time, the Galactic gas was contaminated by ejecta of
Type Ia  supernovae (exploding white dwarfs). These produce little
of the $\alpha$ elements but large amounts of iron-group nuclei.
Thus, the relative abundance of $\alpha$ elements to iron
declined.

Accurate data on the abundances of $\alpha$ elements for stars
 in the halo and disk
 are essential if one
is to extract information about the nucleosynthesis of these
elements. First estimates of the S abundance in halo stars were
provided by \citet{takeda} and \citet{israel} who found that
[S/Fe] increased approximately linearly with decreasing [Fe/H]:
[S/Fe] $\simeq -0.4\times \mathrm{[Fe/H]}$ from the Sun ([Fe/H]
$\equiv 0$) to [Fe/H] $\simeq -3$ , the metallicity limit of the
samples. In contrast, \citet{nissen:IAU210} provide strong
evidence for a constant [S/Fe]$\simeq 0.35$ for stars with [Fe/H]
$\,{\small \raisebox{-0.02cm}{\begin{minipage}{0.2cm}
\raisebox{-0.2cm}{$< $} \\
\raisebox{0.08cm}{$\sim$ }
\end{minipage}}}\, -1$ , and a quasi-linear rise to this value from [Fe/H] = 0.
This discrepancy among the observational results for [Fe/H]
\,{\small \raisebox{-0.02cm}{\begin{minipage}{0.2cm}
\raisebox{-0.2cm}{$< $} \\
\raisebox{0.08cm}{$\sim$ }
\end{minipage}}}\,$-1$  deserves to be resolved before S abundances are added to the
observational inventory relevant to interpretations of Galactic
chemical evolution. This paper describes our attempt to resolve
the discrepancy.

A problem with obtaining the  sulfur abundance of metal-poor stars
is the paucity of suitable atomic  lines. Candidate lines of
S\,{\sc i} lie in the near-infrared (NIR): multiplets of
high-excitation lines occur near  $8694$\,\AA\  and at
$9212-9238$\,\AA . The drawback with the lines at $8694.0$\,\AA\
and $8694.6$\,\AA\ is their weakness in halo stars. \citet{israel}
and \citet{takeda} observed these lines. On the other hand, a
disadvantage with the much stronger triplet at $9212.9$, $9228.1$,
and $9237.5$~\AA\ is the heavy interference by the telluric H$_2$O
lines. \citet{nissen:IAU210} observed these
$\lambda\lambda9213-38$ lines and corrected for the H$_2$O line
absorption.

In this project, we chose to observe the $\lambda\lambda9213-38$
lines in  a number of the stars observed at
$\lambda\lambda8694-95$\,\AA\ by \citet[][in the following
shortened as TH]{takeda}  and \citet[][abbreviate as
I\&R]{israel}. Our stars sample the metallicity range down to
$\mathrm{[Fe/H]}\simeq-3$, and consist of three from the sample of
I\&R, three from  TH, and four from \citet{francois:88}, whose
equivalent widths were reanalyzed by TH and I\&R. Our primary goal
was to check the high [S/Fe] reported previously from the
$\lambda\lambda8694-95$ lines. This check is executed using high
quality spectra of the $\lambda\lambda9213-38$ lines analyzed
using model atmospheres computed for the same effective
temperature and surface gravity as used in the pioneering studies
from which our star-list is made. As a secondary goal, we sought
to gain an indication of how [S/Fe] behaves among halo stars of
different metallicities.

\section{Observations}

 \begin{figure}
   \centering
   \caption{In the upper panel, the full line displays the continuum-normalized
   spectrum of our giant star HD~2665 (including telluric lines, mostly water vapor lines)
   and, as dots,
   the continuum-normalized spectrum of our rapidly rotating B-star,
   which shows only telluric absorption lines.
   In the lower panel,
   the HD~2665 spectrum is divided by this telluric spectrum,
   resulting in a pure stellar spectrum of HD~2665. The sulfur lines
   and the Paschen $\zeta$ hydrogen line emerge clearly after the
   division. Magnesium and iron lines are also marked. The
   abscissa and ordinate scales are the same in both panels.
   Observe that the local continuum of the HD~2665 spectrum is
   fitted with a low order function, to avoid eliminating the
   wings of the hydrogen line. This will be even more important for
   the dwarfs}
              \label{HD2665_both}%
    \end{figure}

   \begin{table*}
      \caption[]{Investigated stars}
         \label{stars}
     $$
        \begin{array}{llcccccccc}
            \hline
            \noalign{\smallskip}
        \mathrm{Star}    & \mathrm{Spectral} & \mathrm{V} &  T_{\mathrm{eff}}  & \log(g) & \mathrm{[Fe/H]} & \xi_\mathrm{micro} & \xi_\mathrm{macro} & \mathrm{Date\,\,of} & \mathrm{Exposure\,\, time}\\
            & \mathrm{type}^\mathrm{a} & [mag] & \mathrm{[K]}& \mathrm { (cgs)} & &\mathrm{[km\,\, s^{-1}}] &\mathrm{[km\,\, s^{-1}}]  & \mathrm{observations}  & \mathrm{[s}]\\
            \noalign{\smallskip}
            \hline
           \noalign{\smallskip}
            \multicolumn{5}{l}{\textrm {Giants and subgiants}}\\
            \noalign{\smallskip}
           \textrm {HD~2665} & \textrm {G5III} & 7.8 & 4990\pm 
           70 & 2.50\pm 0.2  & -1.74\pm0.15 & 1.5\pm0.5&  5.5\pm0.5 &\textrm{Nov. 29, 2001} & 7,200 \\
           \textrm {HD~88609} & \textrm {G5IIIw} & 8.6 & 4570\pm 100 & 0.75\pm0.15 & -2.85\pm0.15 & 1.9\pm0.5  & 7.0\pm0.5 & \textrm{Apr. 27, 2002} &   3,600 \\
           \textrm {HD~111721} & \textrm {G6V} & 8.0 & 5010\pm 100 & 2.31\pm0.15 & -1.27\pm0.15 & 1.2\pm0.5  & 4.5\pm0.5 & \textrm{Apr. 27, 2002}  & 3,600\\
           \textrm {HD~165195} & \textrm {K3p}  & 7.3 & 4190\pm 100 & 1.00\pm0.15 & -1.75\pm0.15 & 1.3\pm0.5 & 7.5\pm0.5 & \textrm{Apr. 27, 2002}   &  1,800 \\
           \hline
            \noalign{\smallskip}
            \multicolumn{5}{l}{\textrm {Dwarfs}}\\
            \noalign{\smallskip}
          \textrm {HD~19445}  & \textrm {sdG5}  & 8.0 & 5810\pm 150 & 4.46\pm0.2 & -1.90\pm0.15  & 1.5\pm0.5 & 4.5\pm0.5 & \textrm{Nov. 29, 2001} &  10,800 \\
           \textrm {HD~84937}  & \textrm {sdF5}  & 8.3  & 6300\pm 100 & 3.97\pm0.15 & -2.06\pm0.15 & 1.1\pm0.5  & 5.0\pm0.5 & \textrm{Nov. 30, 2001}  &  9,000 \\
           \textrm {HD~94028}  & \textrm {F4V} &  8.2  & 5980\pm 100 & 4.30\pm0.15 & -1.35\pm0.15 & 1.5\pm0.5   & 4.5\pm0.5 & \textrm{Apr. 27, 2002}  & 3,600 \\
           \textrm {HD~132475} & \textrm {F5/F6V}  & 8.6  & 5810\pm 100 & 3.91\pm0.15 & -1.38\pm0.15 & 1.9\pm0.5  & 4.5\pm0.5 &\textrm{Apr. 27, 2002}  &  3,600\\
           \textrm {HD~201891} & \textrm {F8V-VI} & 7.4 & 5880\pm 100 & 4.25\pm0.15 & -1.03\pm0.15   & 1.5\pm0.5   & 4.0\pm0.5 &\textrm{Apr. 27, 2002}  &  1,800 \\
           \textrm {HD~201889} & \textrm {G1V}  & 8.0 & 5615\pm100 & 4.24\pm0.2 & -0.71\pm0.15 & 1.5\pm0.5 & 4.0\pm0.5 & \textrm{Apr. 27, 2002}   & 1,800 \\
            \noalign{\smallskip}
            \hline

         \end{array}
     $$
\begin{list}{}{}
\item[$^{\mathrm{a}}$] From Simbad
(\tt{http://simbad.u-strasbg.fr})
\end{list}
   \end{table*}

The investigated stars, all lying in the northern sky, are
presented in Table \ref{stars}. I\&R determined the sulfur
abundance for 8 new stars, including two upper limits, with the
4.2 m William Herschel Telescope on La Palma. We have analyzed
three of these, namely HD~2665, HD~19445, and HD~201889. TH made
new determinations of the sulfur abundance in 6 stars (with one
upper limit) using the {\sc hires} spectrograph on the Keck
telescope on Hawaii. Of these we have observed three (HD~84937,
HD~88609, and HD~165195). Furthermore, we have observed HD~111721,
HD~94028, HD~132475, and HD~201891, which were originally observed
and analyzed by \citet{francois:88}, and then reanalyzed both by
TH and I\&R.

The stars, which span the metallicity range of $-2.75$ {\small
\raisebox{-0.02cm}{\begin{minipage}{0.2cm}
\raisebox{-0.2cm}{$< $} \\
\raisebox{0.08cm}{$\sim$ }
\end{minipage}} }[Fe/H]\footnote{The bracket notation is defined as follows:
$\mathrm{[A/B]\equiv \log(A/B)_\star-\log(A/B)_\odot}$}{\small
\raisebox{-0.02cm}{\begin{minipage}{0.2cm}
\raisebox{-0.2cm}{$< $} \\
\raisebox{0.08cm}{$\sim$ }
\end{minipage}} }$-0.7$,
were observed in observing runs from November 29 and 30, 2001 and
April 27, 2002 at the  W. J. McDonald Observatory using the 2.7-m
Harlan J. Smith telescope with the high-resolution 2dcoud\'e
cross-dispersed echelle spectrograph \citep{tull:95}. The
resolving power was approximately $R= 60\,000$, a value determined
from the {\sc fwhm} of thorium lines in the wavelength calibration
frames. Full spectral coverage from $3400$ to $10\,900$~\AA\ can
be obtained in two exposures. However, we observed every star in
only one setting of the echelle, thus retrieving the entire
wavelength range but with gaps. Our bandpass includes the  sulfur
$\lambda\lambda9213-38$ lines  but not the $\lambda\lambda8694-95$
lines.
We used the TK3 detector, a
$2048 \times 2048\, \mathrm{pixel}^2$ CCD with $24\mu$m pixels.
The observing times, ranging from a half-an-hour to 3 hours for
the F, G, and K stars, are given in Table \ref{stars}. To avoid a
large number of cosmic hits, individual exposures were limited to
a maximum of 30 minutes.

The observed CCD data was processed with the reduction package
{\tt IRAF} to retrieve one-dimensional, continuum normalized, and
wavelength calibrated, pure stellar spectra.
In the wavelength calibration a
root-mean-square of the fit of less than $4$ m\AA\ is achieved.
 Multiple exposures of the same star were added in the
form of 1-dimensional spectra by weighting the different spectra
by their mean count levels.
When several frames are available a cosmic-ray-hit
rejection algorithm was applied.

The local continua of the F, G, and K star spectra were fitted and
normalized by a low-order Legendre function in order not to remove
the wings of the stellar Paschen line. As can be seen in Figure
\ref{HD2665_both}, one of the sulfur lines lies in the wing of
this hydrogen line. The hydrogen lines in the dwarf spectra have
wings stretching over a considerable wavelength range and we
therefore have to proceed with caution when fitting the continuum.

The $9200\,$\AA\ region of interest here is marred by numerous
telluric absorption lines (Figure \ref{HD2665_both}). To eliminate
the telluric contribution to the spectra, we observed
fast-rotating-B-stars with
$\mathrm{v\,sin}i=150-350\,\mbox{km\,s$^{-1}$}$. The intrinsic,
B-star spectra are featureless except for a more or less
pronounced, rotationally broadened Paschen $\zeta$ H {\sc i} line
at $9229.02$~\AA , which means that they can be used as a template
for the telluric absorption lines. For these B-star spectra, we
used a spline function of high order to fit and normalize the
spectra, but also to eliminate the broad Paschen $\zeta$ H{\sc i}
feature and occurring fringes. The hot stars were observed close
in time and at a similar airmass as the programme stars, and were
required to have a signal-to-noise ratio of several hundred. To
achieve this in the 9200~\AA\ region, exposure times from tens of
seconds up to an hour were required, depending on the brightness
of the individual star and the specific spectral type. The reduced
B-star spectra achieved in this manner are assumed to contain
purely telluric absorption lines. The programme spectra were
subsequently divided by a scaled version of the `telluric'
spectra. The divided, reduced spectra are shown in Figures
\ref{HDall_giant} and \ref{HDall_dwarf}. As can be seen in these
Figures, the telluric lines are astonishingly well eliminated
while the stellar sulfur lines show up clearly. While our
procedure of eliminating the telluric lines adds to the noise, the
resultant signal-to-noise ratio of the ratioed stellar-spectra is
satisfactory for our purpose.

The equivalent widths of the sulfur lines at $9212.9$~\AA\ and
$9237.5$~\AA , which are considered to be unblended, were measured
by Gaussian fits or by total integration of the line. The
equivalent width of the $9228.1$~\AA\ line is not measured (except
for HD 111721) since it lies in the stellar Paschen $\zeta$ H{\sc
i} line. The measured equivalent widths of the two (three) sulfur
lines, the equivalent widths of the S{\sc i} $8694.6$~\AA\ line
supplied in the literature (I\&R; TH), and our measured equivalent
widths of six Fe\,{\sc ii} lines (lying at $4416.8$, $5264.8$,
$5284.1$, $6149.3$, $6432.7$, and $6456.4$\,\AA ) are presented in
Table \ref{eqw}. We note that the measured equivalent widths of
the sulfur lines at $9212.9$, ($9228.1$), and $9237.5$~\AA\  range
from approximately 10 to 100 m\AA , which is suitable for a
reliable abundance analysis. Observe also, that the equivalent
widths are an order of magnitude larger than those determined for
the line at $8694.6$~\AA . The $8694.0$~\AA\ line is a factor of
five weaker still and unmeasurable in all the metal-poor stars we
have been investigating, except HD 201889.

We note that the TK3 detector used consists of a thick chip, which
is not severely affected by fringing. The existing fringing
cancels well using a flat field - the top panel is Figure
\ref{HD2665_both} compares two such flat-fielded spectra. Fringing
is eliminated to a high degree by using a hot star observed at the
same time and the lower panel of Figure \ref{HD2665_both} shows
just such a spectrum.

In Figures \ref{HDall_giant} and \ref{HDall_dwarf}, there exist
low points which originate from the cancellation of the very
strongest telluric lines. None of these affect the SI lines. For
example, in the spectrum of HD 19445 in Figure \ref{HDall_dwarf}
only the 9212.9 \AA\, line lies in a wing of a modestly strong
telluric line. The other ones are free. The abundance derived from
the free 9237.5 \AA\, line is $\log A_\mathrm{S}=5.51$ (the
logarithmic number-abundance of sulfur relative to hydrogen, $\log
A_\mathrm{S}=\log N_\mathrm{S}/N_\mathrm{H}$, normalized on a
scale where $\log A_\mathrm{H} = 12.00$), whereas the ratioed
9212.9 \AA\,line gives 5.49. The synthetic spectrum which fits
these two lines, as well as the 9228.1 \AA\, line is calculated
with an abundance of 5.50. The fact that the three sulfur lines
are synthesised well with one sulfur abundance shows that the
elimination of the telluric lines produces a reliable result,
clearly within the other uncertainties.  We estimate the error in
the equivalent widths to be less than 5\%, implying an uncertainty
in the derived sulfur abundance of the order of 0.05 dex, as
estimated from the individual measurements.

In some regions, for example around 9207 \AA\, in the spectrum of
HD 111721 in Figure \ref{HDall_giant} and around 9245 \AA\, in the
spectrum of HD 132475 in Figure \ref{HDall_dwarf}, the elimination
of the telluric lines is hampered due to closely lying telluric
lines with wings blending with each other. The equivalent widths
of the sulfur lines are, however, measured by drawing a local
continuum which should reduce systematic errors due to this
effect.

\begin{table*}
      \caption[]{Measured equivalent widths given in m\AA }
         \label{eqw}
     $$
         \begin{array}{lcccccccccc}
            \hline
            \noalign{\smallskip}
        \mathrm{Star}     & \textrm{W}_\textrm{S\sc i}  & \textrm{W}_\textrm{S\sc i} & \textrm{W}_\textrm{S\sc i} & \textrm{W}_\textrm{S\sc i} & \textrm{W}_\textrm{Fe\sc ii} & \textrm{W}_\textrm{Fe\sc ii} & \textrm{W}_\textrm{Fe\sc ii} & \textrm{W}_\textrm{Fe\sc ii} & \textrm{W}_\textrm{Fe\sc ii} & \textrm{W}_\textrm{Fe\sc ii} \\
           \noalign{\smallskip}
          &  9212.9\,\mathrm{\AA\ }  &   9228.1\,\mathrm{\AA\ } &   9237.5\,\mathrm{\AA\ } &   8694.6\,\mathrm{\AA\ } &   4416.8\,\mathrm{\AA\ } &   5264.8\,\mathrm{\AA\ } &   5284.1\,\mathrm{\AA\ } &   6149.3\,\mathrm{\AA\ } &   6432.7\,\mathrm{\AA\ }  &   6456.4\,\mathrm{\AA\ }  \\
           \noalign{\smallskip}
              \hline
           \noalign{\smallskip}
            \multicolumn{3}{l}{\textrm {Giants and subgiants}}\\
            \noalign{\smallskip}
           \textrm {HD~2665}  & 45  & - & 22 & 3.0^{\mathrm{a}} & - & 15 &26 & 7.3 & 13 & 25 \\
           \textrm {HD~88609} & 24  & -  & 13  & <2.2^{\mathrm{b}} & 32 & 7.0 & 13 & 2.7 & 5.7 & 11 \\
           \textrm {HD~111721} & 78 & 61.0  & 54 & 4.5^{\mathrm{c}} & 73 & 28 & 40 & 16 & 24 & 41\\
           \textrm {HD~165195}  & 40 & -  & 28 & 2.0^{\mathrm{b}} & 65 & 20 & 36 & 9.0 & 18 & 29\\
           \hline
            \noalign{\smallskip}
            \multicolumn{3}{l}{\textrm {Dwarfs}}\\
            \noalign{\smallskip}
           \textrm {HD~19445}  & 30  & -  &  15 & 2.7^{\mathrm{a}} & 13 & 2.0 & 4.5 & 1.3 & 1.5 & 5.7 \\
           \textrm {HD~84937}   & 38 & -  & 14 & 3.4^{\mathrm{b}}& 14 & 1.8 & 3.7 & 1.2 & 2.1 & 5.8 \\
           \textrm {HD~94028}   & 64  & -  & 37 & 4.5^{\mathrm{c}}& 33 & 7.2 & 13 & 5.4 & 8.5 & 18 \\
           \textrm {HD~132475}  & 67 & -   & 40 & 4.2^{\mathrm{c}} & 37 & 9.1 & 17 & 6.7 & 9.8 & 26\\
           \textrm {HD~201891}  & 77 & -  & 52 & 7.1^{\mathrm{c}} & 46 & 14 & 23 & 11 & 12 & 31\\
           \textrm {HD~201889}  & 104  & -  & 70 & 14^{\mathrm{a}}  & 54 & 27 & 35 & 17 & 19 & 42\\
            \noalign{\smallskip}
            \hline
           \noalign{\smallskip}

         \end{array}
     $$

\begin{list}{}{}
\item[$^{\mathrm{a}}$] From Israelian \& Rebolo (2001)

\item[$^{\mathrm{b}}$] From Takada-Hidai et al. (2002)

\item[$^{\mathrm{c}}$] From Fran\c{c}ois (1988)

\end{list}

  \end{table*}

\section{Analysis}

We  analyze our data by modelling the stellar atmosphere and
 requiring that the measured
equivalent widths are reproduced for the two unblended S {\sc i}
$\lambda9213$ and $\lambda9238$ lines. The mean sulfur abundance
yielded in this way is subsequently used in calculating a
synthetic spectrum for a given atmosphere of the entire region
($9200-9500$~\AA ). The synthetic spectrum is thereafter convolved
with a macroturbulence function in order to fit the shapes and
widths of the lines, see Figures \ref{HDall_giant} and
\ref{HDall_dwarf}. In this section we will discuss the model
atmospheres, the stellar parameters including uncertainties, the
line data, and the spectrum synthesis.

 \begin{figure*}
   \centering
   \caption{Continuum-normalized spectra of our programme giants: HD~2665,
   HD~88609, HD~111721, and HD~165195. Telluric absorption lines are ratioed out,
 but some
   residual signatures are  visible.
 In the spectra of HD~2665 and HD~111721 several
   Fe\,{\sc i} lines and a Mg\,{\sc ii} line are discernable and in the spectrum of
   HD~165195 the most conspicuous Fe\,{\sc i} line is detected. In the lower right panel in
   Figure \ref{HDall_dwarf} identifications of all observable metal lines in this
   wavelength region can be found. Our best model spectra are also
   plotted with full lines. See text for a comment on the bad fit of the hydrogen Paschen line
   in the spectrum of HD~165195. After the names the star-model's temperature, $\log g$, metallicity, and microturbulence
   are indicated}
              \label{HDall_giant}%
    \end{figure*}
\begin{figure*}
   \centering
   \caption{Continuum normalized spectra of our programme dwarfs: HD~19445, HD~84937,
   HD~94028, HD~132475, HD~201891,  and HD~201889. The telluric absorption lines are
   removed as well as possible, but some residuals  remain, most prominently in the
spectrum of
   HD~19445 and HD~84937. Metal lines have been identified in the spectra of the four
   lowest panels. In the spectrum of the most metal-rich star,
   HD~201889, all the identified  Fe {\sc i}, Mg {\sc ii}, and S {\sc
   i} lines are marked. The full lines represent our modelled spectra.
   After the names the star-model's temperature, $\log g$, metallicity, and microturbulence
   are indicated
   }

              \label{HDall_dwarf}%
    \end{figure*}

\subsection{Model atmospheres}

I\&R and TH used {\sc atlas9} model atmospheres whereas we use
model atmospheres provided by  the {\sc marcs} code. The {\sc
marcs} code was first developed by \citet{marcs:75} and has been
successively updated ever since. These hydrostatic, plane-parallel
model photospheres are computed on the assumptions of Local
Thermodynamic Equilibrium (LTE), homogeneity and the conservation
of the total flux (radiative plus convective; the convective flux
being computed using the mixing length formulation). Data on
absorption by atomic species are collected from the {\sc vald}
database \citep{VALD} and Kurucz (1995, private communication).
Absorption by molecules is included but quite unimportant for our
stars.

We have computed model atmospheres in plane-parallel geometry
which is an excellent approximation for the dwarfs. To test the
validity of the approximation for the giant stars we calculated a
model atmosphere in spherical geometry for the giant HD~88609 and
compared it with a plane-parallel model. The equivalent widths of
the sulfur lines are changed by only a few percent. We have,
therefore, calculated model atmospheres for all our stars in
plane-parallel geometry.

In our model atmospheres the enhancement of  C, O, Ne, Mg, Si, S,
Ar, Ca, and Ti (that is  the $\alpha$ elements) was assumed to be
$[\alpha/\mathrm{Fe}]=0.4$ except for HD~201889 for which it was
assumed to be $[\alpha/\mathrm{Fe}]=0.2$, due to its higher
metallicity. The derived S abundance is quite insensitive to the
$\alpha$ enhancement.

Using their equivalent widths, line parameters and set of
fundamental stellar parameters, we are able to reproduce the
derived abundances in I\&R and TH. Our use of {\sc marcs} model
atmospheres lowers the abundances by a few hundredths of a dex
relative to those obtained from {\sc atlas9} models used
previously. Clearly, the choice of {\sc atlas9} vs. {\sc marcs} is
not an important factor affecting the derived abundances.

\subsection{Stellar parameters}

We have chosen the same stellar, fundamental parameters for the
model atmospheres (actually $T_{\mathrm{eff}}$,  $\log(g)$, and
$\xi_\mathrm{micro}$) as those determined in I\&R and TH. However,
the metallicities of the stars ($\mathrm{[Fe/H]}$) we redetermined
from six singly ionized iron lines.
 The fundamental stellar
parameters including our determined metallicities and are given in
Table \ref{stars}.

As was noted above, the stars HD~111721, HD~94028, HD~132475, and
HD~201891, which were originally observed and analyzed by
\citet{francois:88}, have been reanalyzed both by  TH and I\&R.
For the determination of the effective temperatures and surface
gravity,  TH took into account interstellar reddening which
resulted in noticeable adjustments of the parameters. We will use
parameters determined by TH for the \citet{francois:88} stars.

TH determined the effective temperatures of HD~84937, HD~88609,
HD~165195, and HD~111721 from the \citet{alonso:99a} calibration,
based on the infrared flux method ({\sc irfm}), also taking the
interstellar reddening into account. For the reanalysis of the
dwarfs HD~94028, HD~132475, and HD~201891, TH used the empirical
temperature scale for dwarfs as formulated by \citet{alonso:96}.
Our new metallicities do not alter this temperature determination
significantly.

Furthermore, TH calculated the surface gravity using
\emph{Hipparcos} parallaxes.
 The microturbulence was determined for
HD~84937, HD~88609, and HD~165195 from Fe {\sc i} lines.
Microturbulence  for HD~111721 is taken from \citet{ryan}, and for
the dwarfs HD~94028, HD~132475, and HD~201891 it was calculated
from the empirical formula provided by \cite{BDP:93}.

For the giant star HD~2665, the effective temperature was
determined by I\&R based on the {\sc irfm} in \citet{alonso:99b}
and the surface gravity ($\log g$) from a non-LTE study of iron
(for references see I\&R). For HD~19445 and HD~201889 the
temperature is based on the {\sc irfm} \citep{israel:98}, and
$\log g$ from an non-LTE analysis of iron by \citet{thevenin}. The
microturbulence we have assumed to be
$\chi_\mathrm{micro}=(1.5\pm0.5)\,\mbox{km\,s$^{-1}$}$ for these
three stars.

\subsection{Line data}

The high-excitation ($\chi=6.5$ eV) S {\sc i} lines used in this
paper are due to  the multiplet $4 s\,^5S^0-4 p^5P$, while the
lines at $8694.0$ and $8694.6$~\AA\ are from the multiplet $4
p\,^5P-4 d^5D^0$. The line parameters (wavelength, excitation
energy, line strength) of our S\,{\sc i} lines and six Fe {\sc ii}
lines are provided in Table \ref{line_parameters}.


In the following,  we have used a solar sulfur abundance of
$\log\varepsilon_\odot(\mathrm S)=7.20$ \citep{chen} as a
reference point. \nocite{thoren}



\begin{table}
      \caption[]{Line parameters}
         \label{line_parameters}
     $$
         \begin{array}{rrrrr}
            \hline
            \noalign{\smallskip}
        \mathrm{Wavelength }     & \textrm{E$_\mathrm{exc}$} & \textrm{$\log\,gf$}  \\ 
            \noalign{\smallskip}
           \textrm{(\AA )} & \textrm{(eV)} &   \mathrm { (cgs)} \\
           \noalign{\smallskip}
              \hline
           \noalign{\smallskip}
            \multicolumn{3}{l}{\textrm {S \sc{i} lines}}\\
            \noalign{\smallskip}

  9212.863  &   6.525  &    0.43^{\mathrm{a}}   &   \\ 
  9228.093  &   6.525  &    0.25^{\mathrm{a}}  &   \\ 
  9237.538  &   6.525  &    0.04^{\mathrm{a}}  &    \\ 
          \hline
            \noalign{\smallskip}
            \multicolumn{3}{l}{\textrm {Fe \sc{ii} lines}}\\
            \noalign{\smallskip}

  4416.830  &  2.778  &  -2.80^{\mathrm{b}}  &  \\ 
  5264.812  &  3.230  &  -3.19^{\mathrm{b}}  &   \\ 
  5284.109  &  2.891  &  -3.29^{\mathrm{b}}  &   \\ 
  6149.258  &  3.889  &  -2.82^{\mathrm{b}}  &    \\ 
  6432.680  &  2.891  &  -3.67^{\mathrm{c}}  &   \\ 
  6456.383  &  3.903  &  -2.25^{\mathrm{c}}  &   \\ 

            \noalign{\smallskip}
            \hline
           \noalign{\smallskip}

         \end{array}
     $$
\begin{list}{}{}
\item[$^{\mathrm{a}}$] Line data taken from the {\sc nist}
database \item[$^{\mathrm{b}}$] From Thor\'{e}n, Edvardsson, \&
Gustafsson, in preparation

 \item[$^{\mathrm{c}}$] From Thor\'{e}n \& Feltzing (2000)

\end{list}
   \end{table}

\subsection{Spectral synthesis}

The sulfur abundance is originally found by reproducing the
equivalent widths of the  unblended sulfur $\lambda9213$ and
$\lambda9238$ lines. This is done with the radiative transfer
routines of the {\sc marcs} codes and yields a mean sulfur
abundance with a standard deviation of 0.01 to 0.1 dex. (This
gives an assessment of the random measuring uncertainties.)
Synthetic spectra around $9200-9250$\,\AA\  are produced by
computations of the radiative transfer through the model
atmospheres, using the mean sulfur abundance and using our line
data for all three sulfur lines and the Paschen $\zeta$ H{\sc i}
line. We calculate the radiative transfer for points in the
spectrum separated by $\Delta \lambda \sim
0.6\,\mbox{km\,s$^{-1}$}$ (corresponding to a resolution of
$\lambda/\Delta \lambda \sim 460\,000$). We subsequently convolve
the spectra with a macro-broadening function, assuming a
Doppler-shift distribution for both radial and tangential velocity
components as specified by the `radial-tangential' model for the
macroturbulence, for details see \citet{gray}. The {\sc fwhm}
velocities for the macroturbulence broadening are given in Table
\ref{stars} and were derived by requiring that the shape and
widths of the sulfur lines should match the observations.

Note, that the sulfur $\lambda9228$ line, which lies in the
Paschen $\zeta$ H{\sc i} line, will also be calculated (including
the hydrogen opacity) by the synthetic spectrum program. For a few
of the dwarfs the wings of the Paschen $\zeta$ H{\sc i} line also
interfere, to various extents, with the two sulfur lines which
were assumed to be unblended. This means that the measured
equivalent widths could be wrong in these cases, underestimating
the sulfur abundance. Therefore, the sulfur abundances were
adjusted in order that the synthetic spectrum, including the
hydrogen-line wings, should fit the observations. The size of
these changes were, however, not more than approximately 0.05 dex
in the sulfur abundance.

In Figures \ref{HDall_giant} and \ref{HDall_dwarf} we have plotted
the final synthetic spectra together with our observations. The
modelled spectra are shifted according to the observed radial
velocities of the stars. From the Figures, we see that all sulfur
lines are synthesized convincingly. Thus, the information from all
three sulfur lines has been used in determining the sulfur
abundance. The synthetic spectra only take the sulfur lines and
the hydrogen line into account. The other metal lines in the
region were not included in the calculation.

The calculations of the wings of the hydrogen line yield a
satisfactory match to the observations, which can be seen, in
particular, in the dwarf spectra (see Figure \ref{HDall_dwarf}).
Recall that the continuum normalization was made with a low order
function only, in order to fit the continuum over a limited range,
so spurious features (such as possible residual fringing) are not
taken out. The hydrogen line cores are expected to be subject to
departures from LTE, which could lead to poor fits. The
unsatisfactory fit of the hydrogen-line core in the giant
HD~165195, for example, could be improved but would require higher
temperatures, of the order of 400 K.






\subsection{Uncertainties in the stellar parameters}

The uncertainties in the fundamental parameters are adopted from
their sources: for  HD~2665, HD~19445, and HD~201889 from I\&R;
for HD~84937, HD~88609, and HD~165195 from TH; and for HD~111721,
HD~94028, HD~132475, and HD~201891 from the reanalysis of the
\citet{francois:88} data by
 TH, see Table \ref{stars}. For HD~2665, HD~19445 and
HD~201889, we assess the uncertainties in the microturbulence to
be of the same order as estimated in  TH. Furthermore, we judge
the uncertainties in the macroturbulence to be of the same order,
that is $\pm0.5\,\mbox{km\,s$^{-1}$}$. The uncertainties in our
new determination of the metallicity is assessed to be less than
$\pm0.15$ dex.

The propagation of these uncertainties into the determination of
the sulfur abundance is calculated by changing the fundamental
parameters of the model atmosphere with the estimated
uncertainties and then running the synthetic spectrum program to
obtain the change in sulfur abundance for a given equivalent
width. Table \ref{error1} shows the consequence of the
uncertainties in the effective temperature, surface gravity,
metallicity, and microturbulence on the [S/H] and [S/Fe] ratios
for one giant and one dwarf. For the coolest stars (e.g.
HD~165195), the dominant source of uncertainty is the 100 K
estimate for the temperature error. The uncertainties in the
sulfur abundance, due to temperature uncertainties, are of the
order of 0.04 dex or less for the dwarfs, and for the giants,
typically 0.1~dex.


\begin{table}
  \caption{Effects on logarithmic abundances derived when changing the fundamental
  parameters of the model atmospheres. Two stars of different parameters are presented.
  The parameters ($\mathrm{T_{eff}/ \log g/[Fe/H]/\xi_t}$) are, respectively:
  $\mathrm{4190/1.0/-1.75/1.3}$) and $\mathrm{6300/3.97/-2.06/1.1}$)}
  \label{error1}
  \begin{tabular}{l l l l l l l l}
  \hline
  \noalign{\smallskip}
    Star & Uncertainty & $\mathrm { \Delta [S/H] }$ &  $\mathrm {\Delta [S/Fe] }$ &   &\\
  \noalign{\smallskip}
  \hline
  \noalign{\smallskip}
 \textrm {HD~165195}& $\delta T_\mathrm{eff}=+100\, \mathrm { K }$ &  $-0.17$ & $-0.08$  \\
 \noalign{\smallskip}
 &  $ \delta \log g =+0.2\, \mathrm { (cgs)}  $ & $+0.10$ & $+0.01$  \\
  \noalign{\smallskip}
 & $\mathrm { \delta [Fe/H]=+0.15  }$ &  $+0.06$ & $+0.01$ \\
  \noalign{\smallskip}
 & $\mathrm { \delta \xi_{\rm micro}=+0.5  }$ &  $-0.12$ & $-0.03$ \\
  \noalign{\bigskip}
 \textrm {HD~84937} & $\delta T_\mathrm{eff}=+100\, \mathrm { K }$ &   $-0.03$ & $-0.04$   \\
 \noalign{\smallskip}
 &  $ \delta \log g =+0.2 \, \mathrm { (cgs)} $ &  $+0.06$ & $-0.01$ \\
  \noalign{\smallskip}
 & $\mathrm {\delta [Fe/H]=+0.15}  $ &  $\pm0.00$ & $\pm0.00$ \\
  \noalign{\smallskip}
 & $\mathrm { \delta \xi_{\rm micro}=+0.5  }$ &  $-0.06$ & $-0.05$ \\
  \noalign{\smallskip}
  \hline
  \end{tabular}
\end{table}


The total uncertainty in the [S/Fe] due to uncertainties in the
fundamental parameters is, in general, of the order of 0.05, but
could be as high as 0.08 in special cases. This is the same
conclusion as \citet{nissen:IAU210} arrive at in their
investigation of the sulfur abundance in solar-type stars.






\subsection{Sulfur and iron abundances}

  \begin{table}
      \caption[]{Abundances of Fe {\sc ii} and S {\sc i}, the latter based on the $\lambda\lambda9213-38$
lines
 }
         \label{9000}
     $$
         \begin{array}{lccccc}
            \hline
            \noalign{\smallskip}
        \mathrm{Star}  &   \log A_\textrm{Fe\sc{ii}} & \textrm{[Fe\sc{ii}/H]}  & \log A_\mathrm{S} & \mathrm{[S/H]} & \mathrm{[S/Fe]} \\
            \noalign{\smallskip}
              \hline
           \noalign{\smallskip}
            \multicolumn{5}{l}{\textrm {Giants and subgiants}}\\
            \noalign{\smallskip}
           \textrm {HD~2665} &  5.76
           & -1.74 & 5.60
           & -1.60 & 0.14  \\
           \textrm {HD~88609} & 4.65
           & -2.85 & 4.74
           & -2.46 & 0.39  \\
           \textrm {HD~111721} & 6.23
           & -1.27 & 6.25
           & -0.95 & 0.32 \\
           \textrm {HD~165195} & 5.75
           & -1.75 & 5.97
           & -1.23 & 0.52 \\
           \hline
            \noalign{\smallskip}
            \multicolumn{5}{l}{\textrm {Dwarfs}}\\
            \noalign{\smallskip}
          \textrm {HD~19445} & 5.60
          & -1.90   &  5.50
          & -1.70 & 0.20 \\
           \textrm {HD~84937} & 5.44
           & -2.06  & 5.27
           & -1.93 & 0.13  \\
           \textrm {HD~94028} & 6.15
           & -1.35  &  5.93
           & -1.27 & 0.08 \\
           \textrm {HD~132475} & 6.12
           & -1.38 &  5.97
           & -1.23 & 0.15 \\
           \textrm {HD~201891} & 6.47
           & -1.03 &  6.26
           & -0.94 & 0.09 \\
           \textrm {HD~201889} & 6.79
           & -0.71  &  6.69
           & -0.51 & 0.20 \\
            \noalign{\smallskip}
            \hline
           \noalign{\smallskip}

         \end{array}
     $$

   \end{table}

The resulting sulfur abundances for our stars are presented in
Table \ref{9000} as $\log A_\mathrm{S}$,
the logarithmic abundance relative to the solar value of $\log
A_{\mathrm{S},\odot} = 7.20$, i.e. $\mathrm{[S/H]}$, and the
logarithmic abundance relative to the solar value normalized to
the metallicity ($\mathrm{[S/Fe]}$). Also, displayed in the Table
are the iron abundances calculated from the measured equivalent
widths of Fe {\sc ii} lines shown in Table \ref{eqw}. The total
uncertainty, including effects of statistical
measuring-uncertainties and uncertainties in the model parameters,
is estimated to be on the order of $\pm0.15$ dex.

  \begin{table}
      \caption[]{Sulfur abundances compared to those from the literature which are based on the $\lambda\lambda8694-95$ lines }
         \label{lit}
     $$
         \begin{array}{lccc}
            \hline
            \noalign{\smallskip}
        \mathrm{Star}     & \log A_\mathrm{S}  &  \log A_\mathrm{S} &  \mathrm{reference}\\
            \noalign{\smallskip}
          & \textrm{$\lambda\lambda9213-38$ lines} & \textrm{$\lambda\lambda8694-95$ lines}   \\
           \noalign{\smallskip}
           & \textrm{this work} & \textrm{literature} &   \\
           \noalign{\smallskip}
              \hline
           \noalign{\smallskip}
            \multicolumn{3}{l}{\textrm {Giants and subgiants}}\\
            \noalign{\smallskip}
           \textrm {HD~2665}  & 5.60   & 5.89  & (1) \\
           \textrm {HD~88609} & 4.74  & <5.49  & (2)  \\
           \textrm {HD~111721} & 6.25  & 6.3/\textbf{6.02}^{\mathrm{a}}/-^{\mathrm{b}} &  (3)  \\
           \textrm {HD~165195}  & 5.97 & 5.98  & (2)  \\
           \hline
            \noalign{\smallskip}
            \multicolumn{3}{l}{\textrm {Dwarfs}}\\
            \noalign{\smallskip}
           \textrm {HD~19445}  &  5.50 & 5.97 & (1) \\
           \textrm {HD~84937}   & 5.27 & 5.70  & (2) \\
           \textrm {HD~94028}   &  5.93 & 6.3/\textbf{6.06}^{\mathrm{a}}/6.09^{\mathrm{b}}  &  (3) \\
           \textrm {HD~132475}  &  5.97 & 6.4/\textbf{6.00}^{\mathrm{a}}/6.08^{\mathrm{b}} &  (3) \\
           \textrm {HD~201891}  &  6.26 & 6.6/\textbf{6.29}^{\mathrm{a}}/6.33^{\mathrm{b}} &  (3) \\
           \textrm {HD~201889}  &  6.69 &  6.83   & (1)  \\
            \noalign{\smallskip}
            \hline
          \noalign{\smallskip}
           \multicolumn{4}{l}{\textrm {(1): \citet{israel}}}\\
           \multicolumn{4}{l}{\textrm {(2): \citet{takeda}}}\\
           \multicolumn{4}{l}{\textrm {(3): Fran\c{c}ois (1988)}}\\
            \noalign{\smallskip}

         \end{array}
     $$

\begin{list}{}{}
\item[$^{\mathrm{a}}$] A reanalysis by Takada-Hidai et al. (2002)
using Fran\c{c}ois's (1988) equivalent widths
\item[$^{\mathrm{b}}$] The corresponding reanalysis by Israelian
\& Rebolo (2001)
\end{list}
   \end{table}

In their non-LTE analysis of neutral sulfur in environments
relevant also in our study, TH find inconsequential effects for
the $\lambda\lambda8694-95$ lines, ranging from 0.00 to 0.08 dex
in the sulfur abundance. \citet{nissen:submitted} find in their
study of the sulfur abundance for similar stars, that the sulfur
abundance they derive from the $\lambda\lambda8694-95$ lines match
the abundance derived from the $\lambda\lambda9213-38$ lines very
well, indicating that departures from LTE should be small also for
the latter lines. It should be noted that neutral sulfur atoms
represent the main ionization state throughout most of the
atmosphere, and most importantly in the line-forming regions.
Thus, we judge that the NLTE effects should be small, justifying a
standard LTE analysis. Furthermore, \citet{nissen:IAU210}
investigate and find that the 3D effects\footnote{By `3D effects'
is meant differences in, for example, derived abundances when
using atmospheric models including 3D, hydrodynamic modelling of
the convection compared with traditional atmospheric models, which
describe convection by the mixing-length approximation.}  for
dwarfs will not severely alter the [S/Fe] ratio (less than 0.05
dex). Therefore, we will neither make any corrections for non-LTE
effects nor for 3D effects. Our $\mathrm{[S/Fe]}$ values are
plotted versus $\textrm{[Fe\sc{ii}/H]}$ in  Figure
 \ref{sulfur_9000_2}. It is clear that  departures from
LTE and  the 3D effects have very similar (small) effects on the
abundance derived from the $\lambda\lambda$9213 -- 38 lines and
that from the $\lambda\lambda8694-95$ lines.

   \begin{figure}
   \centering
   \caption{This figure shows the Galactic chemical evolution of
   sulfur.
   The  iron and sulfur abundances,
   as we have determined it from analyzing Fe {\sc ii} lines and sulfur lines in the NIR, are
   presented with star symbols. Typical errorbars for our measurements of $\pm 0.15$
   dex, both for [S/Fe] and [Fe/H], are indicated. Circles indicate the sulfur and
   iron abundances derived by
   \citet{israel} and squares display values found by \citet{takeda} (including an upper limit
   which is
   illustrated with a triangle). For the 6 star we have in common with \citet{israel} and
   \citet{takeda}, we have connected our new determinations with their measurements, which
   are based on the sulfur $\lambda\lambda8694-95$ lines. We
   have also plotted
   sulfur abundances derived by \citet{nissen:IAU210} and \citet{chen} for southern stars. These are plotted with
   crosses}
              \label{sulfur_9000_2}%
    \end{figure}

Since sulfur abundances have been obtained previously for each of
our stars using the $\lambda\lambda8694-95$ lines, it is of
considerable interest to compare S (and Fe) abundances. This
comparison is made in  Tables \ref{lit} and \ref{lit2}.

  \begin{table*}
      \caption[]{The iron and sulfur abundance ratios ([Fe/H], [S/H], and [S/Fe]) from our work and from the
      literature for the $\lambda\lambda8694-95$ lines}
         \label{lit2}
     $$
         \begin{array}{lccccccc}
            \hline
            \noalign{\smallskip}
        \mathrm{Star}     & \textrm{[Fe/H]} & \textrm{[S/H]} & \textrm{[S/Fe]} & \textrm{[Fe/H]} & \textrm{[S/H]} & \textrm{[S/Fe]} & \mathrm{reference}\\
            \noalign{\vspace{-5pt}}
 &\multicolumn{3}{c}{\underbrace{\hspace{100pt}}} & \multicolumn{3}{c}{\underbrace{\hspace{220pt}}} \\
 \noalign{\smallskip}
         & \multicolumn{3}{c}{ \textrm{based on the $\lambda\lambda9213-38$ lines}} & \multicolumn{3}{c}{ \textrm{study based on the $\lambda\lambda8694-95$ lines} } &  \\
           \noalign{\smallskip}
              \hline
           \noalign{\smallskip}
            \multicolumn{3}{l}{\textrm {Giants and subgiants}}\\
            \noalign{\smallskip}
           \textrm {HD~2665}  & -1.74 & -1.60 & 0.14  & -2.0 & -1.31\pm0.14 & 0.69 & (1) \\
           \textrm {HD~88609} &-2.85 & -2.46 & 0.39 & -2.73 & <-1.72 & <0.99 & (2)  \\
           \textrm {HD~111721} & -1.27& -0.95 & 0.32 & -1.57/\textbf{-1.29}^{\mathrm{a}}/-  & -0.86/\textbf{-1.19}^{\mathrm{a}}/- & 0.71/\textbf{0.07}\pm \textbf{0.41}^{\mathrm{a}}/-& (3)  \\
           \textrm {HD~165195}  & -1.75& -1.23 & 0.52 & -1.87 & -1.23 & 0.64\pm0.09 & (2)  \\
           \hline
            \noalign{\smallskip}
            \multicolumn{3}{l}{\textrm {Dwarfs}}\\
            \noalign{\smallskip}
           \textrm {HD~19445}  & -1.90& -1.70 & 0.20 & -1.88 & -1.23\pm0.12 & 0.65& (1) \\
           \textrm {HD~84937}   & -2.06 & -1.93 & 0.13  & -2.11 & -1.51 & 0.60\pm 0.06& (2) \\
           \textrm {HD~94028}   & -1.35 &  -1.27 & 0.08 & -1.48/\textbf{-1.26}^{\mathrm{a}}/-1.31^{\mathrm{b}}  & -0.91/\textbf{-1.15}^{\mathrm{a}}/-1.12^{\mathrm{b}} & 0.57/\textbf{0.11}^{\mathrm{a}}/0.19^{\mathrm{b}}& (3) \\
           \textrm {HD~132475}  & -1.38 &  -1.23 & 0.15 & -1.37/\textbf{-1.23}^{\mathrm{a}}/-1.37^{\mathrm{b}} & -0.81/\textbf{-1.21}^{\mathrm{a}}/-1.13^{\mathrm{b}} & 0.56/\textbf{0.01}^{\mathrm{a}}/0.24^{\mathrm{b}}& (3) \\
           \textrm {HD~201891}  & -1.03 & -0.94 & 0.09 & -1.25/\textbf{-0.98}^{\mathrm{a}}/-0.87^{\mathrm{b}} & -0.61/\textbf{-0.92}^{\mathrm{a}}/-0.88^{\mathrm{b}} & 0.64/\textbf{0.05}^{\mathrm{a}}/-0.01^{\mathrm{b}} & (3) \\
           \textrm {HD~201889}  & -0.71 & -0.51 & 0.20  & -0.94 & -0.37\pm0.1& 0.57 & (1)  \\
            \noalign{\smallskip}
            \hline
          \noalign{\smallskip}
           \multicolumn{4}{l}{\textrm {(1): \citet{israel}}}\\
           \multicolumn{4}{l}{\textrm {(2): \citet{takeda}}}\\
           \multicolumn{4}{l}{\textrm {(3): Fran\c{c}ois (1988)}}\\
            \noalign{\smallskip}

         \end{array}
     $$

\begin{list}{}{}
\item[$^{\mathrm{a}}$] A reanalysis by Takada-Hidai et al. (2002)
using Fran\c{c}ois's (1988) equivalent widths
\item[$^{\mathrm{b}}$] The corresponding reanalysis by Israelian
\& Rebolo (2001)
\end{list}
   \end{table*}

Three stars -- one giant and two dwarfs -- are in common with
I\&R. Our sulfur abundances are systematically smaller:
$\Delta\log A_S = \log A_S(\rm{ours}) - \log A_S(\rm{theirs})
 = -0.29$ (HD~2665), $-0.47$ (HD~19445), and $-0.14$ (HD~201889).
Recall that we have used the same stellar parameters. It is likely
to be significant that the agreement is best for the star
(HD~201889) for which the $\lambda\lambda8694-95$ lines are
strongest. The difference is in fact $-0.14$ dex also when the
abundance of I\&R is corrected to the {\sc nist} gf-value
(abundance increased by 0.02 dex) and to the {\sc marcs} model
(abundance decreased by 0.02 dex). We suppose that the larger
differences for the two more metal-poor stars reflect the greater
uncertainty in measuring very weak  S\,{\sc i} $\lambda8695$ line
(equivalent widths of about 3 m\AA). Alternative choices of model
atmosphere parameters cannot erase the abundance difference
resulting from the choice of S\,{\sc i} lines.
 The
corresponding $\Delta$s for [Fe/H] are $+0.26$ (HD~2665), $-0.02$
(HD~19445), and $+0.23$ (HD~201889). I\&R indicate that the iron
abundance determination is
 based on Fe\,{\sc i} lines and a non-LTE analysis.
We greatly prefer to derive the Fe abundance from Fe\,{\sc ii}
lines, for which non-LTE effects are slight, than to use Fe\,{\sc
i} lines and an uncertain non-LTE correction. Fe\,{\sc ii} lines
are, however, more sensitive than Fe\,{\sc i} lines to
uncertainties in the surface gravity, but this is also true for
S\,{\sc i} lines compared to S\,{\sc ii}. Thus, this sensitivity
should cancel out in the S\,{\sc i}/Fe\,{\sc ii} ratio.

There are large differences in the [S/Fe] ratios obtained by us
and I\&R: 0.14 vs. 0.69 (HD~2665), 0.20 vs. 0.65 (HD~19445), and
0.20 vs. 0.57 (HD~201889) arising from a higher S abundance in
each case and a lower Fe abundance in two cases. In short, we do
not confirm the high [S/Fe] proposed by I\&R.

Three of our stars were observed and analyzed previously by
 TH. For HD~88609, the most metal-poor star ([Fe/H] =
$-2.8$) of the trio,
 their upper limit to the S abundance is a safe
0.7 dex higher than our measured abundance.  For HD~84937 with
[Fe/H] $= -2.1$, the sulfur abundance derived by  TH is 0.45 dex
larger than ours but this difference is probably due to the
uncertain equivalent width of the $\lambda8695$ line; TH note that
their spectrum is of low signal-to-noise ratio and marred by
residual fringing. In the case of HD~165195, our and their S
abundances agree well. These comparisons use  LTE abundances
derived by  TH; the corrections for non-LTE effects are small. The
LTE Fe abundances derived by  TH from a set of Fe\,{\sc ii} lines
are in good agreement with ours: $\Delta$s are $-0.12$, $+0.12$,
and $+0.05$. The [S/Fe] values for the two stars for which  TH
detected the $\lambda8695$ line in the sense ours vs. theirs, are
0.52 vs. 0.64 (HD~165195) and 0.13 vs. 0.60 (HD~84937).
 TH detected the $\lambda\lambda8694-95$ lines in
three stars not observed by us. Their detections appear quite
secure. We note that the [S/Fe] ratio inferred by  TH for these
stars, with [Fe/H] $\simeq -1.4$, average about 0.2 and fit well
with our suggested run of [S/Fe] with [Fe/H], see Figure
\ref{sulfur_9000_2}.

We chose four stars from \citet{francois:88}. This quartet was
reanalysed by  TH and I\&R using equivalent widths of the
$\lambda8695$ line reported by \citet{francois:88}. Table
\ref{lit} shows that, for the dwarfs, these reanalyses and our
abundance from the NIR lines are in good agreement. The agreement
extends to [S/Fe], cf. Table \ref{lit2}. The typical total
uncertainties quoted by  TH, for their dwarfs, for the [S/Fe]
ratio is 0.15 dex. However, for their analysis of HD~111721 the
uncertainty is larger (0.4 dex).


%

%

\section{Discussion}

Our observations show that the S\,{\sc i} NIR lines are detectable
and S abundances determinable to [Fe/H] $\simeq -3$, and, thus,
confirm the conclusions of \citet{nissen:IAU210} from their {\sc
vlt/uves} observations of the same lines in southern dwarfs and
subgiants. In Figure \ref{sulfur_9000_2}, we show our results with
those of \citet{nissen:IAU210} for halo stars, also based on the
$\lambda\lambda$9213 -- 38 lines,
 and \citet{chen} for disk stars. \citet{chen} used the $\lambda\lambda8694-95$
lines and also lines at 6046, 6053, and 6757~\AA , but for their
chosen disk stars these lines provide readily measurable lines in
high-resolution spectra.
In  Figure \ref{sulfur_9000_2}, our values for the stars in common
with I\&R and  TH are connected with theirs to guide the eye. For
the  stars we have in common with  TH, we have, for consistency,
chosen their [Fe{\sc ii}/H]$_\mathrm{LTE}$ and [S/Fe{\sc
ii}]$_\mathrm{LTE}$ values.
The figure illustrates the comparisons between our and the
$\lambda\lambda$8694 -- 95-based abundances discussed in the
previous section.

There is no evidence in our data  for the rise of [S/Fe] with
decreasing [Fe/H] earlier proposed by I\&R and  TH. I\&R claimed a
linear trend such that $\mathrm{[S/Fe]} \simeq 0.9$ at
$\mathrm{[Fe/H]} = -2$, and  TH put the slope lower so that [S/Fe]
= 0.5 at $\mathrm{[Fe/H]} = -2$. We have suggested that the lower
[S/Fe] now being obtained are due to a combination of
overestimates of the strength of the very weak $\lambda\lambda
8694-95$ lines in the most metal-poor stars, and of underestimates
of the Fe abundances by I\&R. We reiterate that the adopted
effective temperatures and surface gravities are those used by
I\&R and  TH. Our primary goal was to check the latter authors' S
abundances,
 as directly as possible, by using $\lambda\lambda$9213 -- 38
lines rather than the $\lambda\lambda$8694 -- 95 lines used by
them.
 Definition of the run of [S/Fe] with [Fe/H] was a secondary goal whose
full achievement demands additional stars.
%
 As can be seen in Figure \ref{sulfur_9000_2}, the
evidence that [S/Fe] is nearly independent of [Fe/H] for [Fe/H]
{\small
\raisebox{-0.02cm}{\begin{minipage}{0.2cm}
\raisebox{-0.2cm}{$< $} \\
\raisebox{0.08cm}{$\sim$ }
\end{minipage}}}$ -1$ is greatly strengthened on adding the sulfur abundances
derived by \citet{nissen:IAU210}, who find a constant value of
$\mathrm{[S/Fe]}\approx0.35$ dex.

There is a hint in Figure \ref{sulfur_9000_2} that our [S/Fe]
estimates may be systematically about 0.1 dex lower than those of
\citet{nissen:IAU210}.
This difference could be the result of an accumulation of minor
differences in effective temperature scales, surface gravity
determinations, Fe\,{\sc ii} gf-values,
and others. We also note, that \citet{nissen:IAU210} derive a
solar iron abundance 0.03 dex larger than the one we apply.
Considering this fact would diminish the discrepancy.

Many studies of abundance of $\alpha$ elements in halo stars have
shown that [$\alpha$/Fe] at a fixed [Fe/H] is largely without
cosmic scatter. The S abundances of \citet{nissen:IAU210} confirm
that this result may now be extended to sulfur.
 In our small sample, HD~165195 appears to
depart from the constant [S/Fe] indicated by other stars. This
departure may possibly be due to adoption of incorrect atmospheric
parameters. We note that  TH remark that the determination of the
effective temperature for HD~165195 is uncertain. They consider a
range of temperatures ($4131-4507$~K) in which their adopted
temperature (and therefore the one used by us) is at the lower end
of this range. Observe, that the sulfur abundances determined for
this star are the most sensitive to the effective temperature of
our entire sample, which can be seen in Table \ref{error1}. A
temperature increase of 300 K would yield a [S/Fe] ratio close to
our mean value. Therefore, our value might be seen as a
determination which could be on the high side. We note, in
passing, that an increase in temperature would also make the
modelled hydrogen-line fit the observed line better.

The exact run of the abundances with metallicity has consequences
for theories and our understanding of Galactic chemical evolution
and for the sites of formation of the elements. So, what are the
sites where sulfur was synthesized during the first few billion
years of the Universe? $\alpha$-capture elements are thought to be
formed during explosions of supernovae (SNe) Type II.  Owing to
the short life-times of high-mass stars, the progenitors of SNe
Type II, the abundances of elements synthesized by them and
ejected into the interstellar medium, quickly reach a steady state
and show a constant over-abundance in the ordinary [S/Fe] vs.
[Fe/H] diagram. However, based on their data,  TH and I\&R discern
a linear rise of the $\alpha$-capture element sulfur for the halo
phase of the Milky Way, thus showing a divergent behaviour
compared to  other $\alpha$-capture elements, except possibly
oxygen. If there is a rise in the abundances with decreasing
metallicity, other processes for their formation have to be
invoked.  TH and I\&R suggest a sulfur contribution from
hypernovae in the early galaxy as an explanation of the high
values. Hypernovae are thought to be the explosion of an extremely
massive star (several 100 M$_\odot$). Another mechanism that could
lead to a divergent evolutionary behaviour of oxygen and sulfur as
compared to the other $\alpha$-capture elements, is the fact that
these two elements are volatile elements as opposed to the others
which are refractory elements. This is important in the modelling
of the transport and mixing of supernovae ejecta into the
interstellar medium, with volatile elements experiencing a mixing
time-scale, more than an order of magnitude faster. However,
considering the results that \citet{nissen:IAU210} and we find,
there does not seem to be a need for a distinct evolutionary
scenario for sulfur as compared to the other $\alpha$ elements.

\section{Conclusions}

We have shown that  reliable abundances of sulfur down to
metallicities of $\mathrm{[Fe/H]}\sim -3$ are obtainable  using
the S {\sc i} lines lying at $9212.9$, $9228.1$, and $9237.5$~\AA
. In an attempt to resolve the discrepancy found in the
literature, about the Galactic chemical evolution of sulfur  for
metallicities lower than $\mathrm{[Fe/H]}${\small
\raisebox{-0.02cm}{\begin{minipage}{0.2cm}
\raisebox{-0.2cm}{$< $} \\
\raisebox{0.08cm}{$\sim$ }
\end{minipage}} }$-1$, we have observed
an ensemble of ten stars, both dwarfs and giants, previously
analyzed with a different and weaker diagnostic; the S {\sc i}
lines at $8694.0$ and $8694.6$\,\AA. These previous investigations
of the 10 stars claimed a rise of the $\mathrm{[S/Fe]}$ ratio with
decreasing metallicity. We are not able to confirm this rise. Our
conclusion is instead that we corroborate the finding by
\citet{nissen:IAU210} indicating a similar behaviour of sulfur to
the other $\alpha$ elements.

We suggest that the reasons for the difference in determined
sulfur abundances and [S/Fe] are the strength of the lines
analyzed and the way the metallicity is determined. First, we have
used lines which are stronger by a factor of ten as compared to
those used earlier. This is critical for metal-poor stars. Thus,
we conclude that the smallest equivalent widths for the weaker
$\lambda8695$ line must have been overestimated in the previous
studies, leading to an overestimation of the [S/H] ratio by up to
$0.5$ dex. Second, we have determined the metallicity from
Fe\,{\sc ii} lines which should give a better determination of the
metallicities than an non-LTE analysis of Fe\,{\sc i} lines.

We also show that in spite of the numerous telluric, water-vapor
lines in the 9200~\AA\  spectral region, a careful reduction of
the data can provide clean spectra. We conclude that the $9212.9$,
$9228.1$, and $9237.5$~\AA\ lines are the preferred ones to be
used for abundance analyses of sulfur of halo stars.










\begin{acknowledgements}
We should like to thank Drs. C. Allende Prieto, B. Gustafsson, K.
Eriksson, B. Edvardsson, A. Kron, and N. Piskunov for valuable
discussions and comments. This research has been supported in part
by the Swedish Research Council, Stiftelsen Blanceflor
Boncompagni-Ludovisi, n\'ee Bildt, the Swedish Foundation for
International Cooperation in Research and Higher Education, and
the Robert A. Welch Foundation of Houston, Texas.
\end{acknowledgements}

\end{document}